%% file: paper-arxiv.tex
\documentclass[10pt,twocolumn]{confpaper}

\graphicspath{ ./ }

\date{}
\title{\ttlfnt{
\TITLE}}
\author{
    \aufnt{Nick Feamster} \\
    \affaddr{Princeton University}
    \and
    \aufnt{Jason Livingood} \\
    \affaddr{Comcast}
}

\begin{document}

\maketitle


\begin{sloppypar}

\input{abstract}
\input{introduction}
\input{background}

\input{limitations2}
\balance\input{future}\label{lastpage}
\end{sloppypar}

\small
\setlength{\parskip}{-1pt}
\setlength{\itemsep}{-1pt}
\footnotesize 
\bibliography{paper}
\bibliographystyle{abbrv}

\end{document}

%% file: abstract.tex
\begin{abstract}
Government organizations, regulators, consumers, Internet service
providers, and application providers alike all have an interest in
measuring user Internet ``speed''. 
Access speeds have
increased by an order of magnitude in past years, 
with gigabit speeds available to tens 
of millions of homes. 
Approaches must evolve to
accurately reflect the changing user experience and network speeds.
This paper offers historical and technical background on current 
speed testing methods, highlights their limitations as access network speeds 
continue to increase, and offers recommendations for the next generation 
of Internet ``speed'' measurement. 
\end{abstract}

%% file: introduction.tex
\section{Introduction}

Various governmental organizations have begun to rely on so-called ``Internet
speed tests'' to measure broadband Internet speed. Examples of these programs
include the Federal Communications Commission's ``Measuring Broadband
America'' program~\cite{fcc-mba}, California's CALSPEED program~\cite{calspeed}, the United
Kingdom's Home Broadband Performance Program~\cite{ofcom}, and various other initiatives in
states including Minnesota \cite{minnestoa}, New
York~\cite{newyork-1,newyork-2,newyork-3}, 
and Pennsylvania~\cite{pennsylvania}.  These programs have
various goals, ranging from assessing whether ISPs are delivering on advertised
speeds to assessing potentially underserved rural areas that could benefit from
broadband infrastructure investments.

The accuracy of measurement is critical to these assessments, as measurements can
inform everything from investment decisions to policy actions and even
litigation.  Unfortunately, these efforts sometimes rely on outmoded
technology, making the resulting data unreliable or misleading. This paper
describes the current state of speed testing tools, outlines their
limitations, and explores paths forward to better inform the various technical
and policy ambitions and outcomes.

Some current speed test tools were well-suited to measuring access link
capacity a decade ago but are no longer useful because they made a design assumption that the
Internet Service Provider (ISP) last mile access network was the most
constrained (bottleneck) link.  This is no longer a good assumption, due to
the significant increases in Internet access speeds due to new technologies.
Ten years ago, a typical ISP in the United States may have delivered tens of
megabits per second (Mbps). Today, it is common to have ten times faster
(hundreds of megabits per second), and gigabit speeds are available to tens of
millions of homes. The performance bottleneck has often shifted
from the ISP access network to a user's device, home WiFi network, network
interconnections, speed testing infrastructure, and other areas. 

A wide range of factors can influence the results of an Internet speed test,
including: user-related considerations, such as the age of the device;
wide-area network considerations, such as interconnect capacity;
test-infrastructure considerations, such as test server capacity; and test
design, such as whether the test runs while the user's access link is
otherwise in use. Additionally, the typical web browser opens multiple
connections in parallel between an end user and the server to increasingly
localized content delivery networks (CDNs), reflecting an evolution of
applications that ultimately effects the user experience.

These developments suggest the need to evolve our understanding of the utility
of existing Internet speed test tools, and consider how these tools may need
to be redesigned to present a more representative measure of a user's Internet
experience.  

%% file: background.tex
\section{Background}\label{sec:background}

In this section, we discuss and define key network performance metrics,
introduce the general principles of Internet ``speed tests'' and explore the
basic challenges facing any speed test.

\subsection{Performance Metrics}

\begin{figure}[t]
\centering\includegraphics[width=0.75\linewidth]{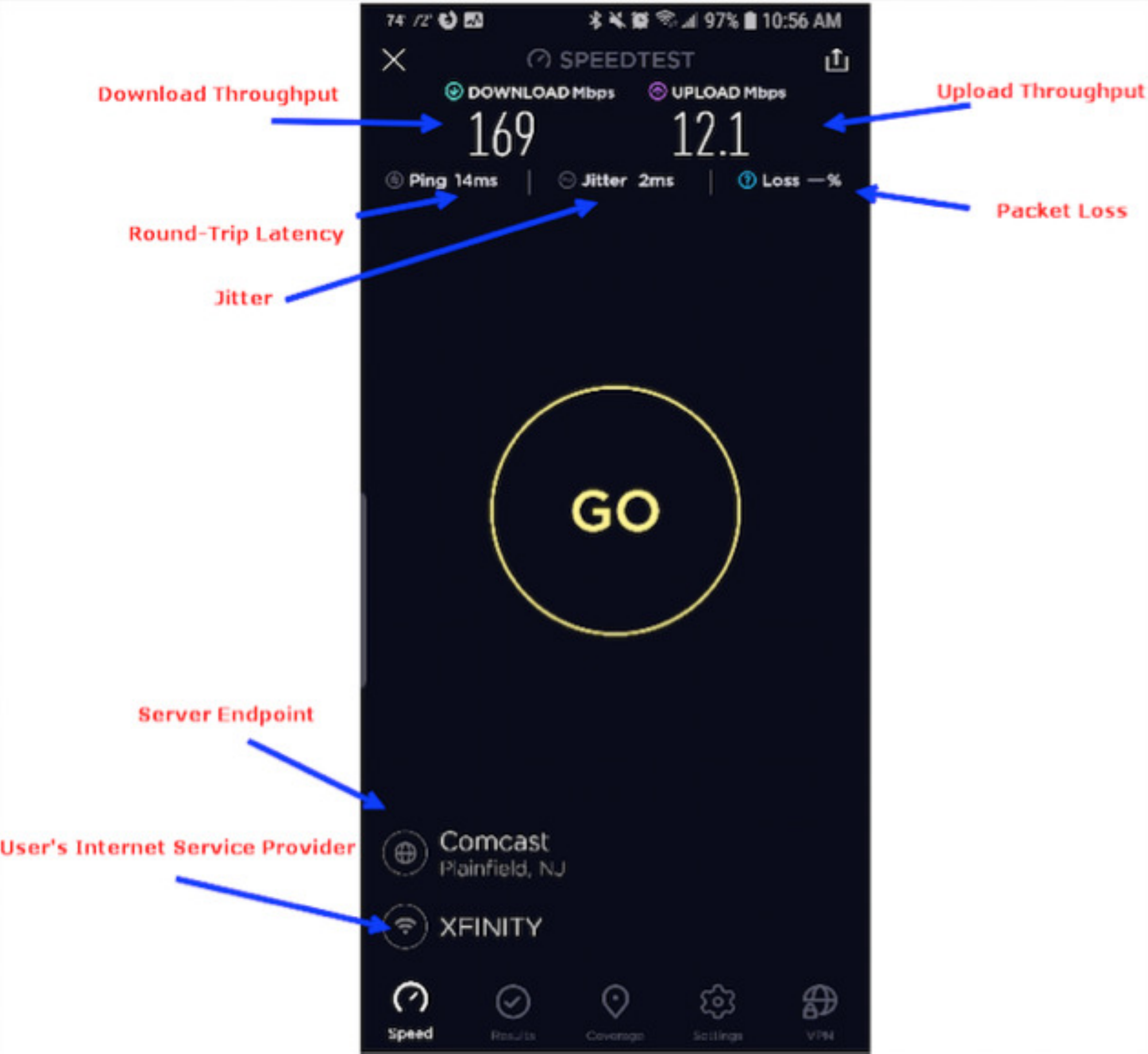}
\caption{Example metrics from an Ookla Speedtest in New Jersey, a canonical
    Internet speed test.}
\label{fig:example-metrics}
\end{figure}

When people talk about Internet ``speed'', they are generally talking about
throughput. End-to-end Internet performance is typically measured with a
collection of metrics---specifically throughput (\ie, ``speed''), latency, and
packet loss.  Figure~\ref{fig:example-metrics} shows an example speed test
from a mobile phone on a home WiFi network. It shows the results of a
``native" speed test from the Ookla Android speed test
application~\cite{ookla-speedtest} run in New
Jersey, a canonical Internet speed test.  This native application reports 
the user’s ISP, the location of the test server
destination, and the following performance metrics:

\paragraph{Throughput} is the amount of data that can be transferred between
two network endpoints over a given time interval. For example, throughput can
be measured between two points in a given ISP's network, or it can be measured
for an end-to-end path, such as between a client device and a server at some
other place on the Internet.  Typically a speed test measures both downstream
(download), from server to client, and upstream (upload), from client to
server (Bauer {\em et al.}~\cite{bauer} offer an in-depth discussion of
throughput metrics). Throughput is not a constant; it changes from minute to
minute based on many factors, including what other users are doing
on the Internet.  Many network performance tests, such as the FCC
test~\cite{fcc-mba} and Ookla's speed test, include additional metrics that
reflect the user's quality of experience.

\paragraph{Latency} is the time it takes for a single data packet to travel to
a destination. Typically latency is measured in terms of {\em round-trip
latency}, since measuring one-way latency would require tight time
synchronization and the ability to instrument both sides of the Internet path.
Latency generally increases with distance, due to factors such as the speed of
light for optical network segments; other factors can influence latency,
including the amount of queueing or buffering along an end-to-end path, as
well as the actual network path that traffic takes from one endpoint to
another.
TCP throughput 
is inversely proportional to end-to-end latency~\cite{tanenbaum1996computer}; all things being equal, then,
a client will see a higher throughput to a nearby server than it will to a
distant one.

\paragraph{Jitter} is the variation between two latency measurements. Large 
jitter measurements are problematic.

\paragraph{Packet Loss Rate} is typically computed as the number of lost
packets divided by the number of packets transmitted. Although high packet
loss rates generally correspond to worse performance, some amount of packet
loss is normal because a TCP sender typically uses packet loss as the feedback
signal to determine the best transmission rate.  Many applications such as
video streaming are designed to adapt well to packet loss without noticeably
affecting the end user experience, so there is no single level of packet loss
that automatically translates to poor application performance. Additionally,
certain network design choices, such as increasing buffer sizes, can reduce
packet loss, but at the expense of latency, leading to a condition known as
``buffer bloat"~\cite{bufferbloat-1,bufferbloat-2}.

\subsection{Speed Test Principles and Best Practices}

\paragraph{Active Measurement.}
Today's speed tests are generally referred to as {\em active measurement
tests}, meaning that they attempt to measure network performance by
introducing new traffic into the network (\ie, so-called ``probe traffic'').
This is in contrast to {\em passive} tests, which observe traffic passing over
a network interface to infer performance metrics. For speed testing, active
measurement is the recognized best practice, but passive measurement can be
used to gauge other performance factors, such as latency, packet loss, video
quality, and so on.

\paragraph{Measuring the Bottleneck Link.}
A typical speed test sends traffic that traverses many network links,
including the WiFi link inside the user's home network, the link from the ISP
device in the home to the ISP network, and the many network level hops between
the ISP and the speed test server, which is often hosted on a network other
than the access ISP.  The throughput measurement that results from such a test
in fact reflects the capacity of the {\em most constrained} link, sometimes
referred to as the ``bottleneck'' link---the link along the end-to-end path
that is the limiting factor in end-to-end throughput.  If a user has a 1~Gbps
connection to the Internet but their home WiFi network is limited to 200~Mbps,
then any speed test from a device on the WiFi network to the Internet will not
exceed 200~Mbps.  Bottlenecks can exist in an ISP access network, in a transit
network between a client and server, in the server or server data-center
network, or other places. In many cases the bottleneck is located somewhere
along the end-to-end path that is not under the ISP's or user's direct
control.

\begin{figure}[t]
\centering\includegraphics[width=0.75\linewidth]{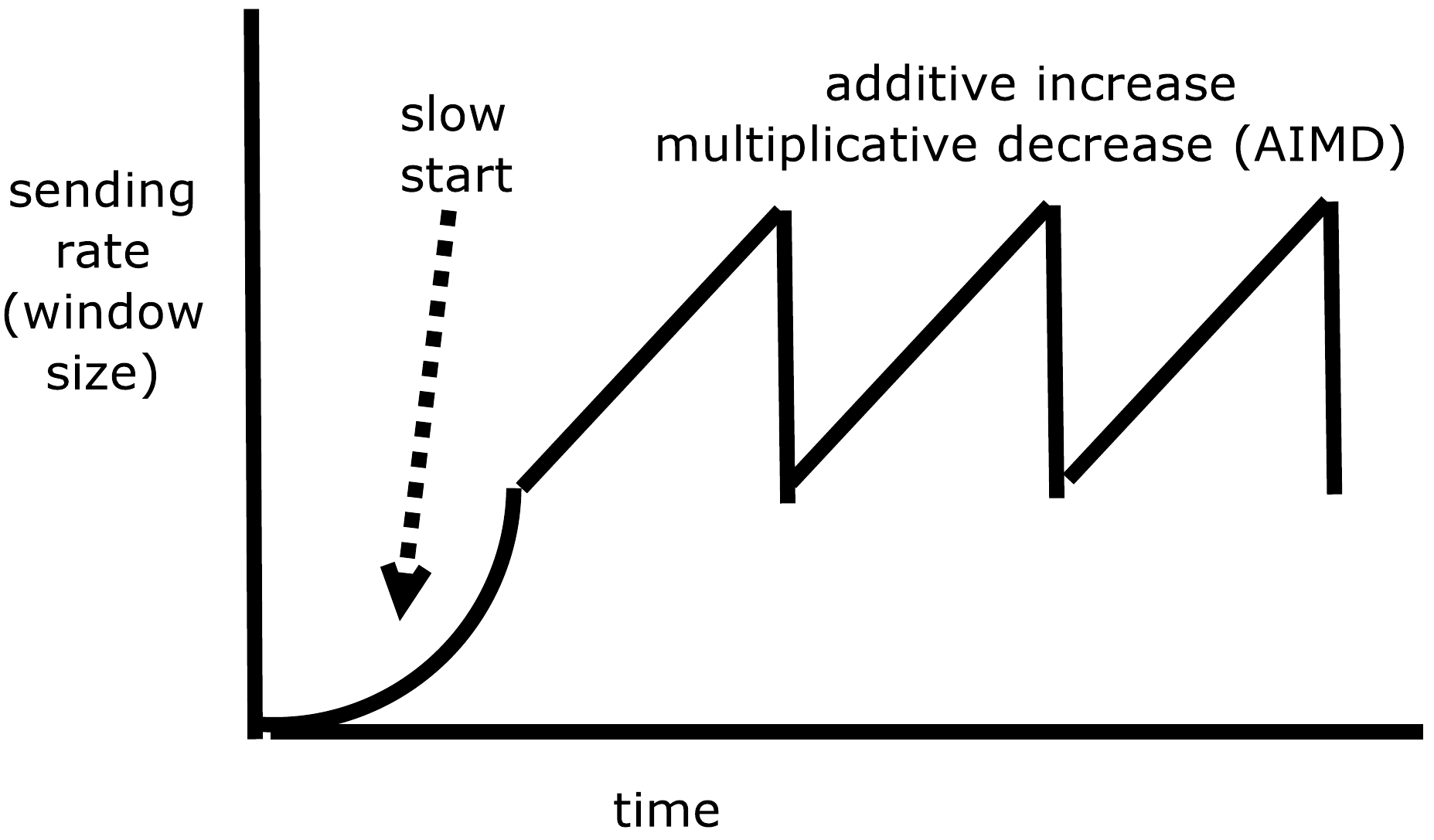}
\caption{TCP Dynamics.}
\label{fig:tcp-slow-start}
\end{figure}

\paragraph{Use of Transmission Control Protocol.}
Speed tests typically use the Transmission Control Protocol (TCP) to measure
throughput.  In keeping with the nature of most Internet application transfers
today---including, most notably, web browsers---most speed tests use multiple
parallel TCP connections.  Understanding TCP's operation is critical to the
design of an accurate speed test.
Any TCP-based speed test should be: (1)~long enough to measure steady-state
transfer; (2)~recognize that TCP transmission rates
naturally vary over time, and (3)~use multiple TCP connections.
Figure~\ref{fig:tcp-slow-start} shows TCP's dynamics, including the initial
slow start phase.  During TCP slow start, the transmission rate is far lower
than the network capacity.  Including this period as part of a throughput
calculation will result in a throughput measurement that is less than the
actual available network capacity. If test duration is too short, the test
will tend to underestimate throughput.  As a result, accurate speed test tools
must account for TCP slow start.  Additionally, instantaneous TCP throughput
continually varies because the sender tries to increase its transfer rate in
an attempt to find and use any spare capacity (a process known as ``additive
increase multiplicative decrease'' or AIMD).  

\begin{figure}[t!]
\begin{minipage}{1\linewidth}
\begin{subfigure}[b]{0.45\linewidth}
\centering\includegraphics[width=0.75\linewidth]{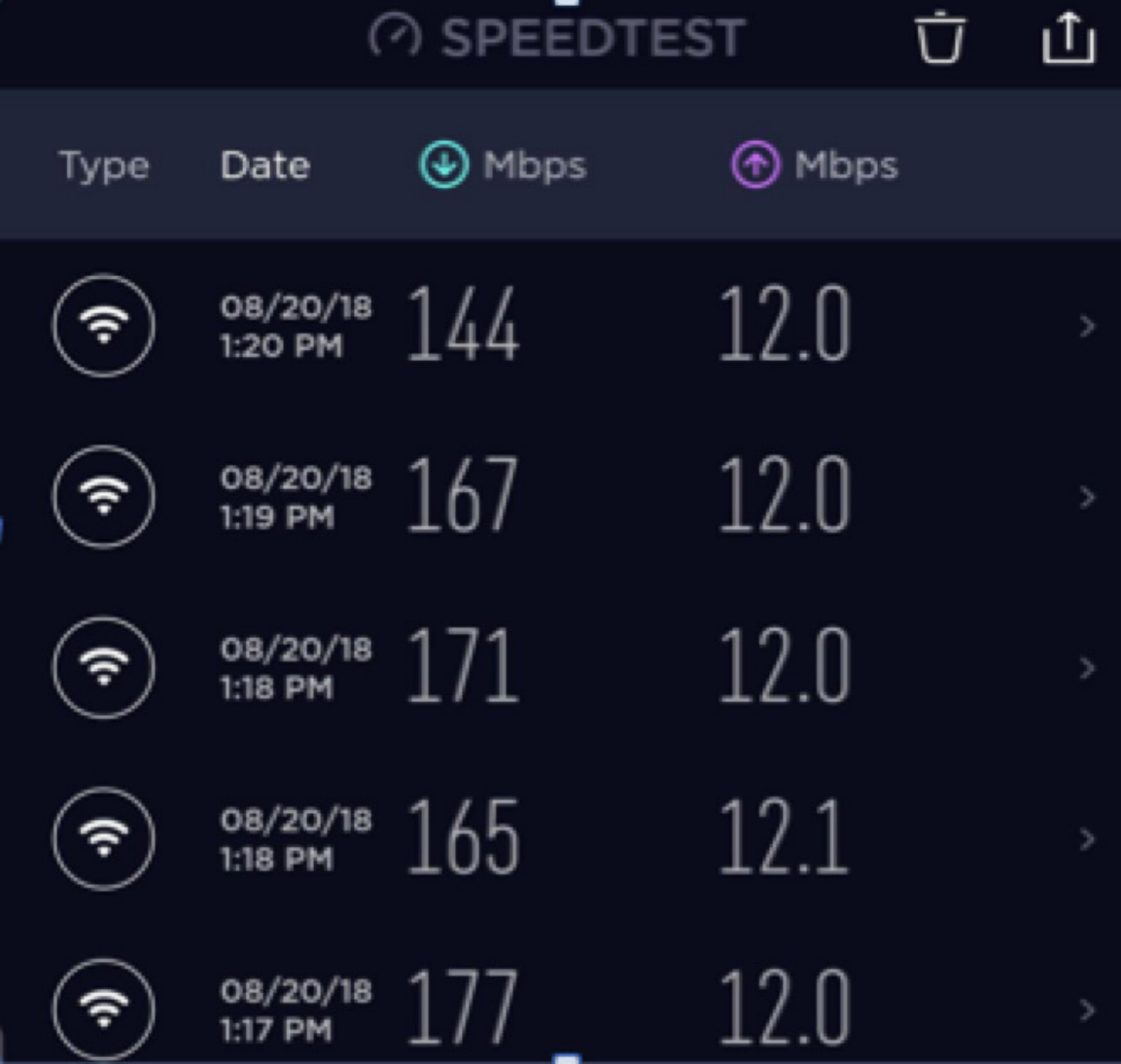}
\caption{Five successive runs of Ookla Speedtest yield variable results on downstream throughput.}
\label{fig:five-ookla-runs}
\end{subfigure} \hfill
\begin{subfigure}[b]{0.45\linewidth}
\centering\includegraphics[width=0.75\linewidth]{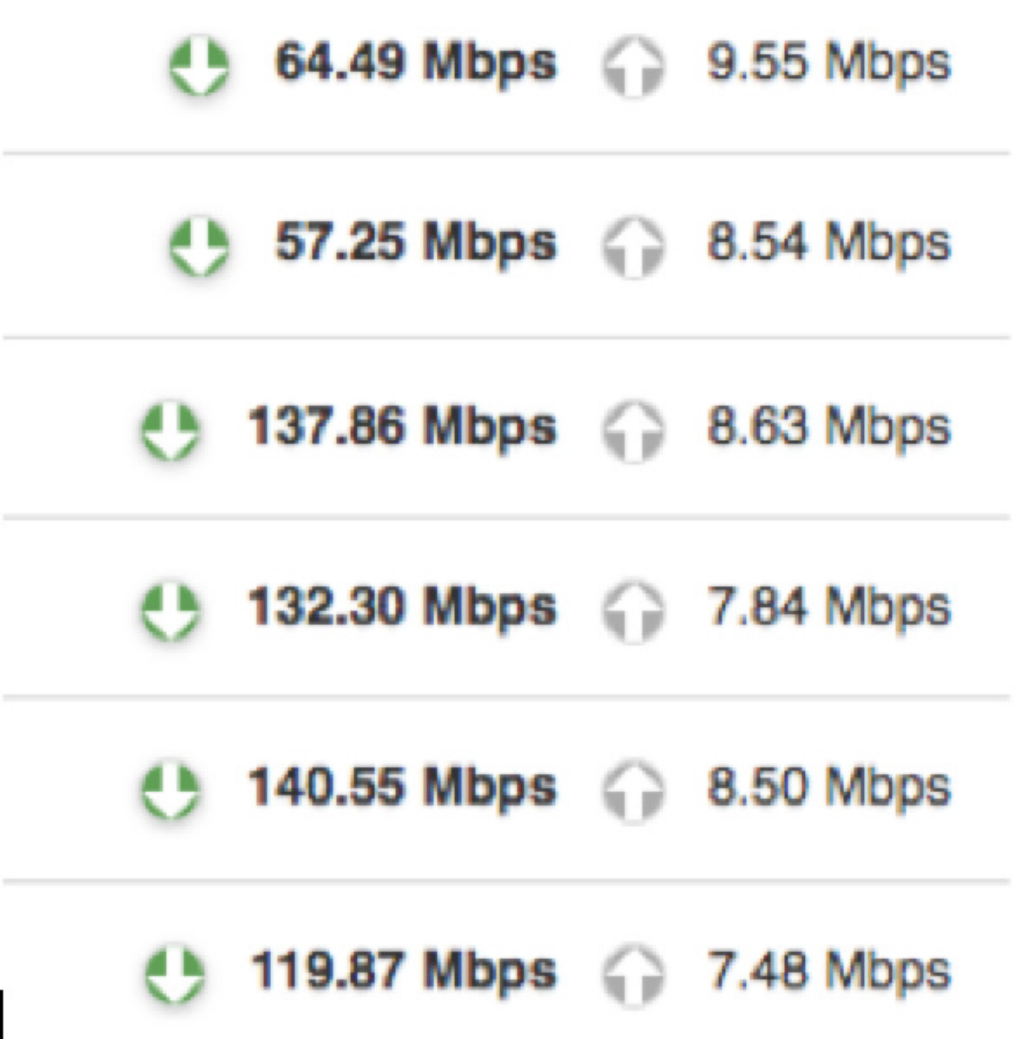}
\caption{Internet Health Test runs in succession to six different
  servers. The test measures consistently lower throughput and also
  shows variability, both to different servers and across successive
  test runs.} 
\label{fig:five-iht-runs}
\end{subfigure}
\end{minipage}
\caption{Successive runs of different throughput tests.}
\label{fig:five-runs}
\end{figure}

\paragraph{Inherent Variability.}
A speed test measurement can produce highly variable results.
Figure~\ref{fig:five-runs} shows an illustrative example of typical
variability that a speed test might yield, both for Internet Health Test (IHT)
and Ookla Speedtest.  These measurements were performed successively on the
same Comcast connection provisioned for 200 Mbps downstream and 10 Mbps
upstream throughput. The tests were performed in succession. Notably,
successive tests yield different measurements. IHT, a web front-end to a tool
called the Network Diagnostic Test (NDT), 
also consistently and significantly under-reports throughput, especially at
higher speeds.

%% file: limitations2.tex
\section{Limitations of Existing Speed Tests}\label{sec:limitations}

Existing speed tests have a number of limitations that have become more acute
in recent years, largely as a result of faster ISP access links and the
proliferation of home wireless networks. The most profound change is that {\em
as network access links have become faster, the network bottleneck has moved
from the ISP access link to elsewhere on the network}. A decade ago, the
network bottleneck was commonly the access ISP link; with faster ISP access
links, the network bottleneck may have moved any number of places, from the
home wireless network to the user's device itself.  Other design factors may
also play a role, including how measurement samples are taken and the
provisioning of the test infrastructure itself.

\subsection{User-Related Considerations}

\paragraph{The home wireless network.} Speed tests that are run over a home
wireless connection often reflect a measurement of the user's home wireless
connection, {\em not} that of the access ISP, because the WiFi network itself
is usually the lowest capacity link between the user and test server\cite{wtf,
eetimes, revolution-wifi,pc-world,lifewire,apple}.  Many factors affect the
performance of the user's home wireless network, including: distance to the
WiFi Access Point (AP) and WiFi signal strength, technical limitation of a
wireless device and/or AP, other users and devices operating on the same
network, interference from nearby APs using the same spectrum, and
interference from non-WiFi household devices that operate on the same spectrum
(\eg, microwave ovens, baby monitors, security cameras).

Many past experiments demonstrate that the user's WiFi---not the ISP---is
often the network performance bottleneck.  
Sundaresan {\em et al.} found that
whenever
downstream throughput exceeded 25 Mbps, the user's home wireless network was
almost always the bottleneck~\cite{wtf}. Although the study is from 2013, and both access
link speeds and wireless network speeds have since increased, the general
trend of home wireless bottlenecks is still prevalent. 

\paragraph{Client hardware and software.} Client types range from dedicated
hardware, to software embedded in a device on the user's network, to native
software made for a particular user operating system, and web browsers. Client
type has an important influence on the test results, because some may be
inherently limited or confounded by user factors. Dedicated hardware examples
include the SamKnows whitebox and RIPE Atlas probe. Embedded software refers to
examples where the software is integrated into an existing network device such
as cable modem, home gateway device, or WiFi access point. A native
application is software made specifically to run on a given operating system
such as Android, iOS, Windows, and Mac OS. Finally, web-based tests simply run
from a web browser. In general, dedicated hardware and embedded software
approaches tend to be able to minimize the effect of user-related factors and
are more accurate as a result.

\begin{figure}[t]
\centering\includegraphics[width=0.75\linewidth]{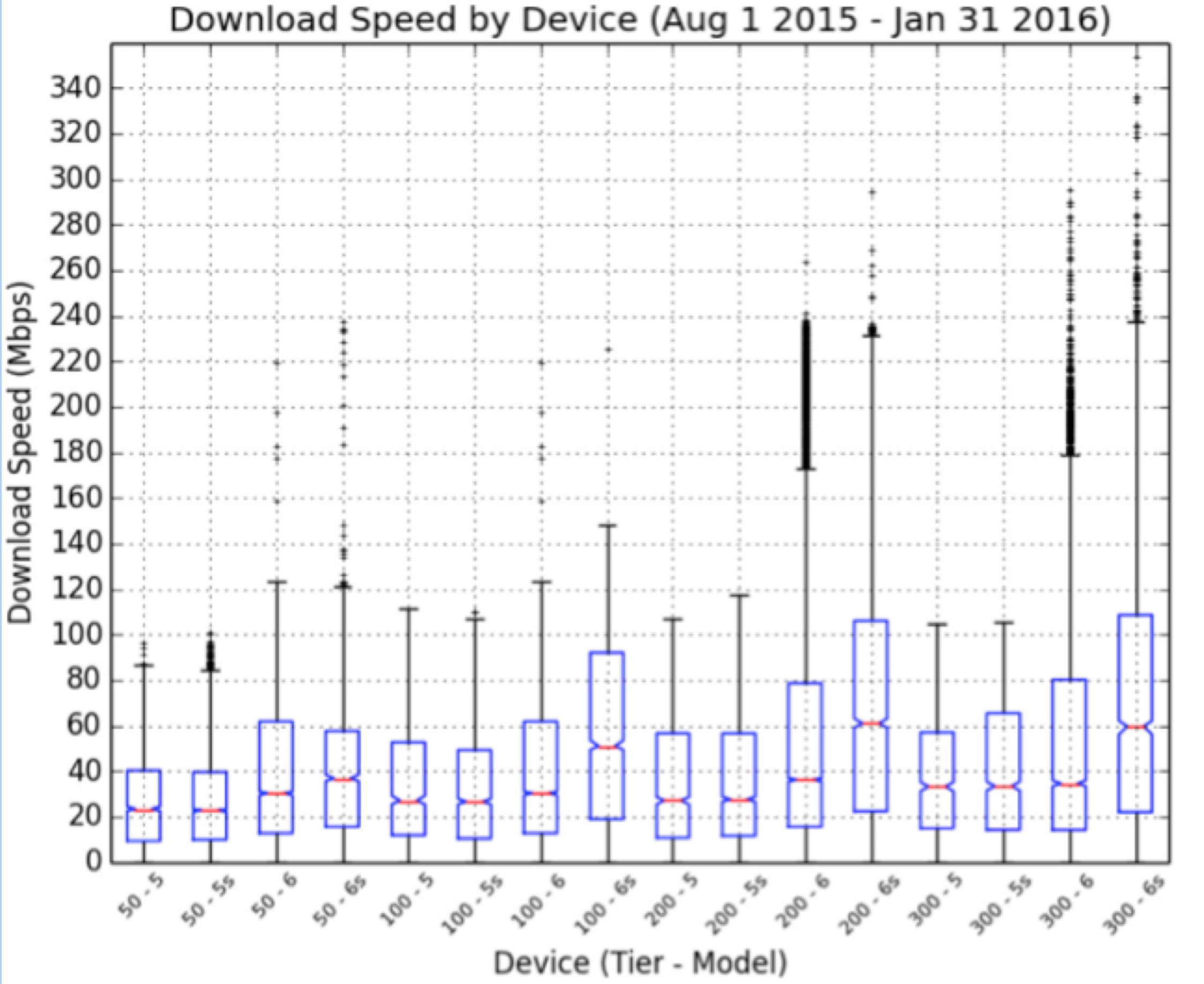}
\caption{Distribution of download speeds across different device types. Older devices
do not support 802.11ac, so fail to consistently hit 100 Mbps.}
\label{fig:dl-by-device}
\end{figure}

Many users continue to use older wireless devices in their homes (\eg, old
iPads and home routers) that do not support higher speeds. 
Factors such as memory, CPU, operating system, and network interface card
(NIC) can significantly affect throughput measurements. For example, if a user
has a 100~Mbps Ethernet card in their PC connected to a 1~Gbps Internet
connection, their speed tests will never exceed 100 Mbps and that test result
cannot be said to represent a capacity issue in the ISP network; it is a
device limitation. As a result, many ISPs document recommended hardware and
software standards \cite{xfinity-requirements}, especially for 1 Gbps
connections.  The limitations of client hardware can be more subtle.
Figure~\ref{fig:dl-by-device} shows an example using iPhone released in
2012--2015.  This shows that any user with an iPhone 5s or older is unlikely
to reach 100 Mbps, likely due to the lack of a newer 802.11ac wireless
interface.

\begin{figure}[t!]
\begin{minipage}{1\linewidth}
\begin{subfigure}[b]{0.45\linewidth}
\centering\includegraphics[width=0.75\linewidth]{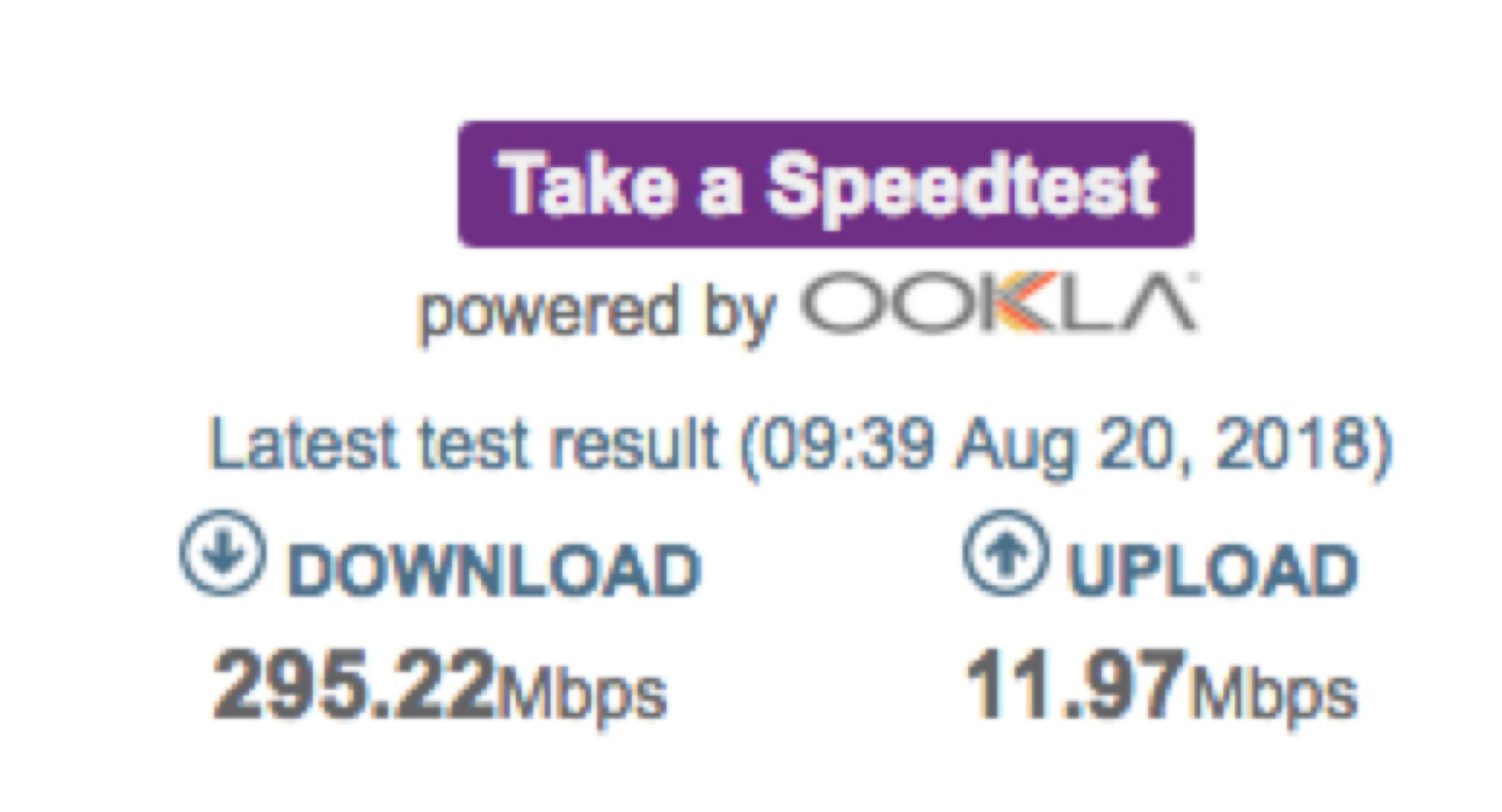}
    \vspace*{0.5in}
\caption{Ookla router-based test.}
\label{fig:ookla-router}
\end{subfigure} \hfill
\begin{subfigure}[b]{0.45\linewidth}
\centering\includegraphics[width=0.75\linewidth]{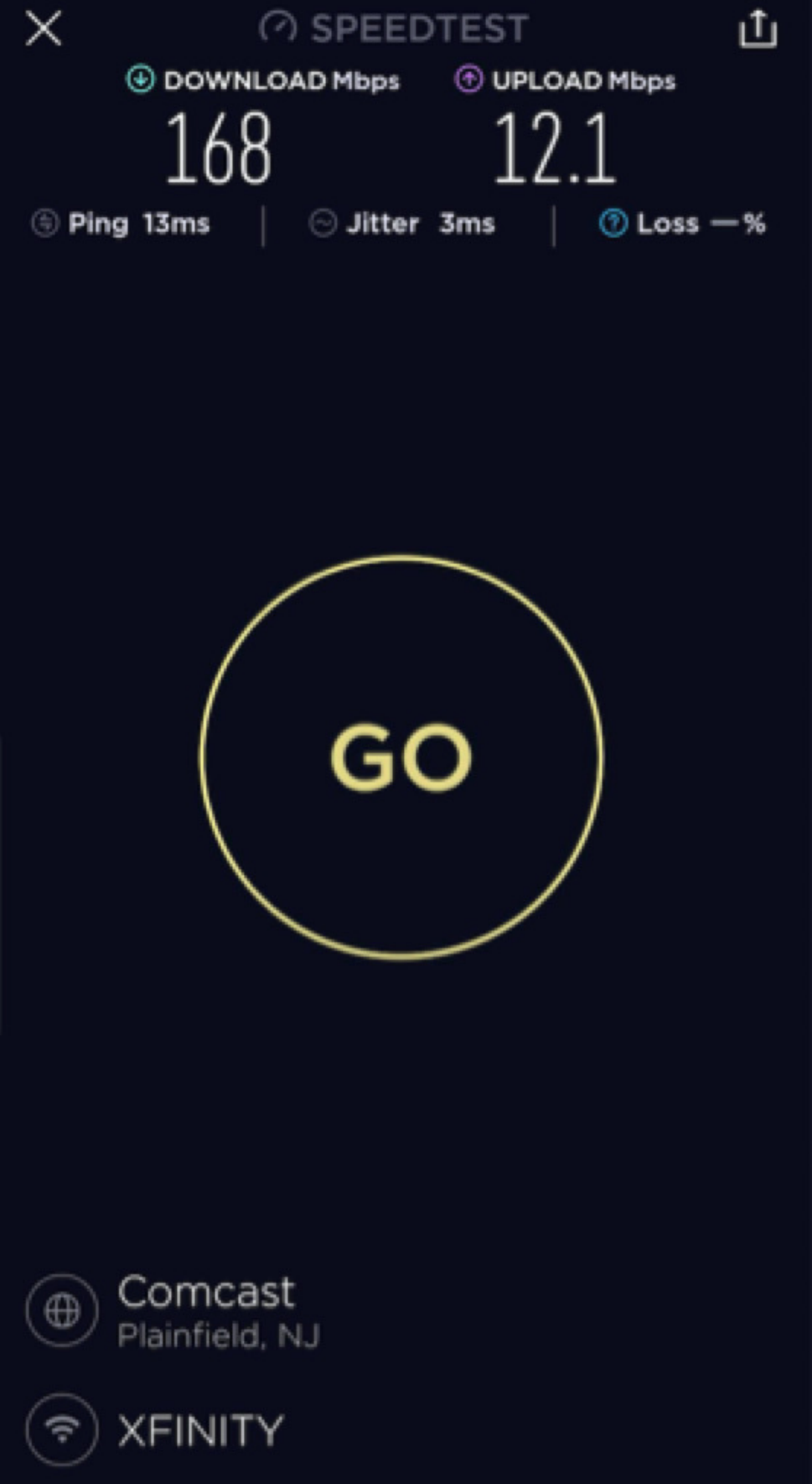}
\caption{Ookla native desktop test.}
\label{fig:ookla-native}
\end{subfigure}
\end{minipage}
\caption{Ookla Speedtest, router-based test and native desktop test from the same home
network.}
\label{fig:ookla-router-native}
\end{figure}

\paragraph{Router-based testing vs. device-based testing.}
Figure~\ref{fig:ookla-router-native} shows an example of two successive speed
tests. Figure~\ref{fig:ookla-router} uses software embedded in the user's router, so that no
other effects of the local network could interfere.
Figure~\ref{fig:ookla-native} shows the same
speed test (\ie, Ookla Speedtest), on the same network, performed immediately
following the router-based test using native software on a mobile device over
WiFi.  The throughput reported from the user's mobile device on the home
network is almost half of the throughput that is reported when the speed test
is taken directly from the router. 

\paragraph{Competing ``cross traffic''.} At any given time, a single network
link is simultaneously carrying traffic from {\em many} senders and receivers.
Thus, any single network transfer must share the available capacity with the
competing traffic from other senders---so-called {\em cross traffic}. Although
sharing capacity is natural for normal application traffic, a speed test that
shares the available capacity with competing cross traffic will naturally
underestimate the total available network capacity.  Client-based speed tests
cannot account for cross traffic;
because the client cannot see the volume of other traffic
on the same network, whereas a test that runs on the user's home router can
account for cross traffic when conducting throughput measurements.

\subsection{Wide-Area Network Considerations}

\paragraph{Impaired ISP Access Network Links} An ISP's ``last mile" access
network links can become impaired.  For example, the quality of a DOCSIS
connection to a home can become impaired by factors such as a squirrel chewing
through a line or a bad ground wire.  Similarly, fixed wireless connections
can be impaired by weather or leaves blocking the antenna. To mitigate the
potential for an individual impairment unduly influencing ISP-wide results,
tests should  be conducted with a large number of users.

\paragraph{Access ISP capacity.} Capacity constraints within
an ISP's network can exist, whether in the access network, regional network
(metropolitan area), or backbone network. Regional and backbone networks
usually have excess capacity so the only periods when they may
be constrained would be the result of a disaster (\eg, hurricane damage) or 
temporary conditions such fiber cuts or BGP hijacking.  Usually ISP capacity
constraints arise in the last-mile access networks, which are by nature shared
in the first mile or first network element, (\eg, passive optical networking
(PON), DOCSIS, DSL, 4G/5G, WiFi, point-to-point wireless). 

\paragraph{Transit and interconnect capacity.}  Another significant
consideration is the connection to ``transit" and ``middle mile" networks. The
interconnects between independently operated networks may also introduce
throughput bottlenecks. As user speeds reach 1~Gbps, ensuring that there are
no capacity constraints on the path between the user and test
server---especially across transit networks---is a major consideration.
In one incident in 2013, a bottleneck in the Cogent transit network reduced
NDT throughput measurements by as much as 90\%.
Test results improved when Cogent began prioritizing NDT
test traffic over other traffic.  Transit-related issues have often affected
speed tests. In the case of the FCC's MBA platform, this prompted them to add
servers on the Level 3 network to isolate the issues experienced with M-Lab's
infrastructure and the Cogent network, and M-Labs has also added additional
transit networks to reduce their reliance on one network. 

\paragraph{Middleboxes.} End-to-end paths often have devices along the
path, called ``middleboxes", which can affect performance.  For example, a
middlebox may perform load balancing or security functions (\eg, malware
detection, firewalls).  As access speeds increase, the capacity of middleboxes
may increasingly be a constraint, which will mean that test results will
reflect the capacity of those middleboxes rather than the access link or other
measurement target. 

\paragraph{Rate-limiting.} Application-layer or destination-based rate limiting,
often referred to as throttling, can also cause the performance that users
experience to diverge from conventional speed tests.  Choffnes {\em et al.}
have developed Wehe, which detects application-layer rate
limiting~\cite{wehe}; thus far, the research has focused on HTTP-based video
streaming de-prioritization and rate-limiting. Such rate limiting could exist
at any point on the network path, though most commonly it may be expected in
an access network or on the destination server network. In the latter case,
virtual servers or other hosted services may be priced by peak bitrate and
therefore a hard-set limit on total peak bitrate or per-user-flow bitrate may
exist. Web software such as Nginx has features for configuring rate
limiting~\cite{nginx-rate}, as cloud-based services may charge by total
network usage or peak usage; for example, Oracle charges for total bandwidth
usage~\cite{oracle-pricing}, and FTP services often enforce per-user and
per-flow rate limits~\cite{ftp-rate}.

\paragraph{Rate-boosting.} Rate-boosting is the opposite of rate limiting; it
can enable a user to temporarily exceed their normal provisioned rate for a
limited period.  For example, a user may have a 100 Mbps plan but may be allowed to
burst to 250 Mbps for limited periods if spare capacity exists. This effect
was noted in the FCC’s first MBA report in 2011 and led to use of a longer
duration test to measure ``sustained'' speeds~\cite{fcc-mba2011}. Such
rate-boosting techniques appear to have fallen out of favor, perhaps partly
due greater access speeds or the introduction of new technologies such as
DOCSIS channel bonding.

\subsection{Test Infrastructure Considerations}

Because speed tests based on active measurements rely on
performing measurements to some Internet endpoint (\ie, a measurement
server), another possible source of a performance bottleneck is the server
infrastructure itself.

\paragraph{Test infrastructure provisioning.} The test server infrastructure
must be adequately provisioned so that it does not become the bottleneck for
the speed tests.  In the past, test servers have been overloaded,
misconfigured, or otherwise not performing as necessary, as has been the case
periodically with M-Lab servers used for both FCC MBA testing and NDT
measurements.  Similarly, the data center switches or other network equipment
to which the servers connect may be experiencing technical problems or be
subject to other performance limitations. In the case of the FCC MBA reports, at one
point this resulted in discarding of data collected from M-Lab servers due to
severe impairments~\cite{fcc-aug-2013,mba-report-2014}. The connection between a given
data-center and the Internet may also be constrained, congested, or otherwise
technically impaired, as was the case when some M-Lab servers were
single-homed to a congested Cogent network. Finally, the servers themselves may
be limited in their capacity: if, for example, a server has a 1~Gbps Ethernet
connection (with real-world throughput below 1 Gbps) then the server cannot be
expected to measure several simultaneous 1 or 2~Gbps tests.  Many other
infrastructure-related factors can affect a speed test, including server
storage input and output limits, available memory and CPU, and so on.
Designing and operating a high scale, reliable, high performance measurement
platform is a difficult task, and as more consumers adopt 1 Gbps services
this may become even more challenging~\cite{apnic-blog}. 


Different speed test infrastructures have different means for incorporating
measurement servers into their infrastructure.  Ookla allows volunteers to run
servers on their own and contribute these servers to the list of possible
servers that users can perform tests against.  Ookla uses empirical
measurements over time to track the performance of individual servers. Those
that perform poorly over time are removed from the set of candidate servers
that a client can use.  Measurement Lab, on the other hand, uses a fixed,
dedicated set of servers as part of a closed system and infrastructure. 
For many years, these servers have been: (1)~constrained by a 1~Gbps uplink;
(2)~shared with other measurement experiments (recently, Measurement Lab has
begun to upgrade to 10~Gbps uplinks). Both of these factors can and
did contribute to the platform introducing its own set of performance
bottlenecks. 

\begin{figure}[t!]
\begin{minipage}{1\linewidth}
\begin{subfigure}[b]{0.45\linewidth}
\centering\includegraphics[width=0.75\linewidth]{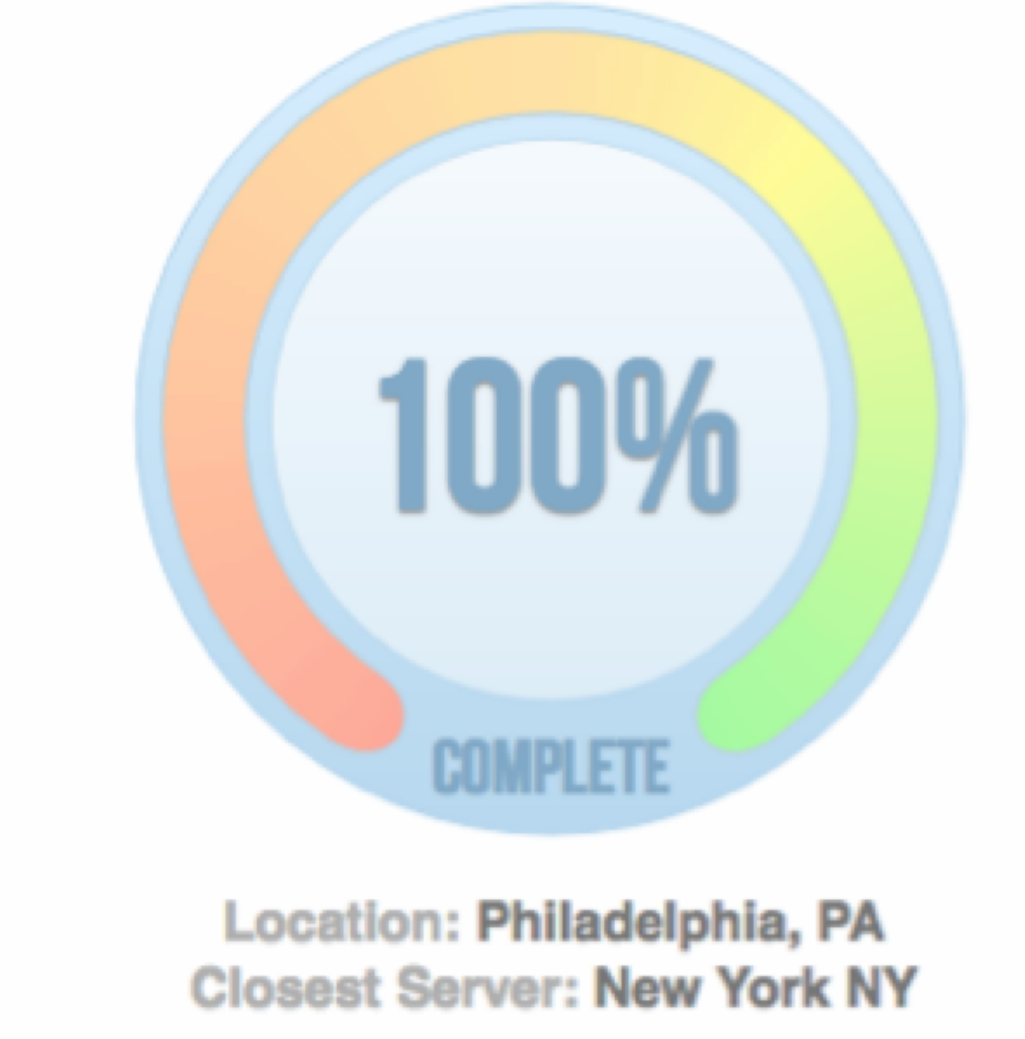}
\caption{Internet Health Test mistakenly locating a client in Princeton, NJ to
    Philadelphia, PA (50+ miles away), and performing a speed test to a server to New York City.}
\label{fig:iht-geo}
\end{subfigure} \hfill
\begin{subfigure}[b]{0.45\linewidth}
\centering\includegraphics[width=0.75\linewidth]{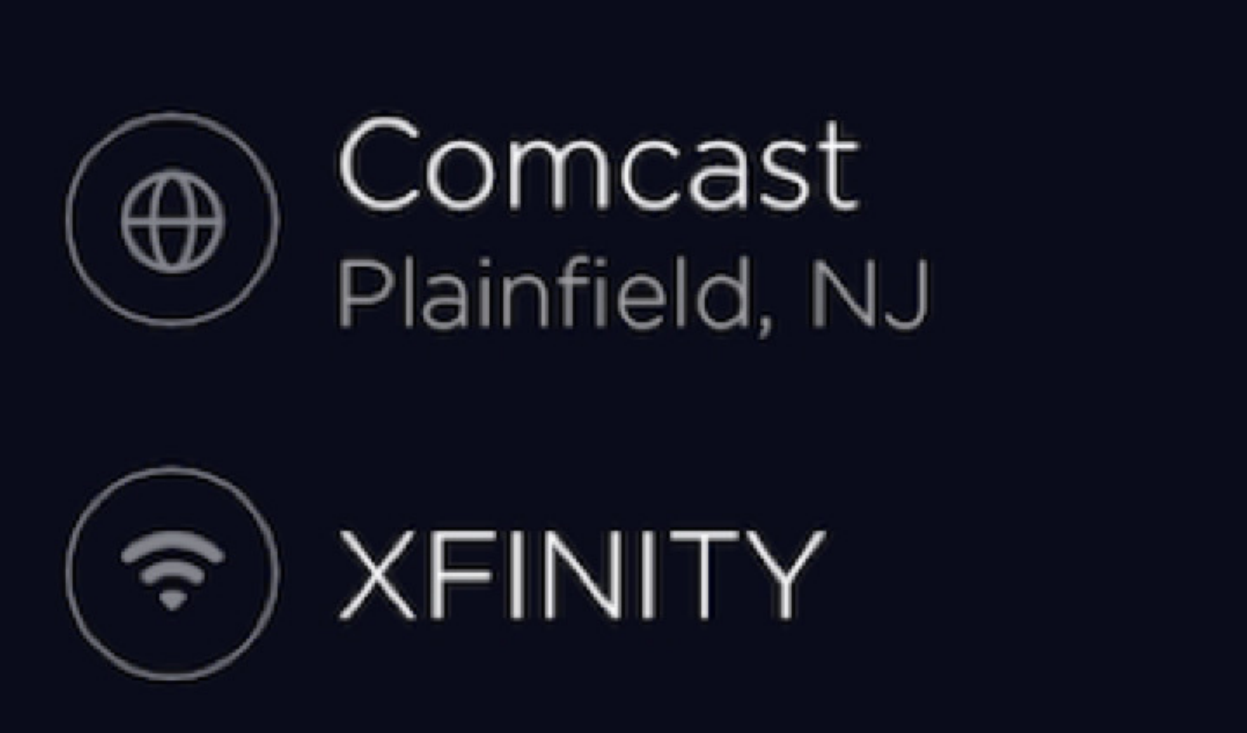}
\caption{Ookla Speedtest directing a client in Princeton, NJ to an on-net Speedtest server in Plainfield, NJ. Ookla also allows a user to select another nearby server.}
\label{fig:ookla-geo}
\end{subfigure}
\end{minipage}
\caption{IHT and Ookla geolocation.}
\label{fig:geo}
\end{figure}

\paragraph{Server placement and selection.} A speed test estimates the
available capacity of the network between the client and the server.
Therefore, the throughput of the test will naturally depend on the distance
between these endpoints as measured by a packet's round trip time (RTT). This
is extremely important, because TCP throughput is inversely proportional to
the RTT between the two endpoints.  For this reason, speed test clients
commonly attempt to find the ``closest" throughput measurement server to
provide the most accurate test result and why many speed tests such as
Ookla's, use thousands of servers distributed around the world.  to select the
closest server, some tests use a process called ``IP geolocation", whereby a
client location is determined from its IP address. Unfortunately, IP
geolocation databases are notoriously inaccurate, and client location can
often be off by thousands of miles. Additionally, latency resulting from
network distance typically exceeds geographic distance, since network paths
between two endpoints can be circuitous, and other factors such as network
congestion on a path can affect latency. Some speed tests mitigate these
effects with additional techniques. For example, Ookla's Speedtest uses IP
geolocation to select an initial set of servers that are likely to be  close,
and then the client selects from that list the one with the lowest RTT (other
factors may also play into selection, such as server network capacity).
Unfortunately, Internet Health Test (which uses NDT) and others rely strictly
on IP geolocation.

Figure~\ref{fig:geo} shows stark differences in server selection
between two tests: Internet Health Test (which relies on IP geolocation and
has a smaller selection of servers); and Ookla Speedtest (which uses a
combination of IP geolocation, GPS-based location from mobile devices, and
RTT-based server selection to a much larger selection of servers).
Notably, the Internet Health Test not only mis-locates the client (determining
that a client in Princeton, New Jersey is in Philadelphia), but it also
selects a server that is in New York City, which is more than 50 miles from
Princeton. In contrast, the Ookla test, which selects an on-network Comcast
server in Plainfield, NJ, which is merely 21 miles away, and also gives the
user the option of using closer servers through the ``Change Server" option.

\subsection{Test Design Considerations}

\paragraph{Number of parallel connections.} A significant consideration in the
design of a speed test is the number of parallel TCP connections that the test
uses to transfer data between the client and server, since the goal of a speed
test is to send as much data as possible and this is usually only possible
with multiple TCP connections.   Using multiple
connections in parallel allows a TCP sender to more quickly and more reliably
achieve the available link capacity. In addition to achieving a higher share
of the available capacity (because the throughput test is effectively sharing
the link with itself), a transfer using multiple connections is more resistant
to network disruptions that may result in the sender re-entering TCP slow
start after a timeout due to lost packets.

\begin{figure}[t]
\centering\includegraphics[width=\linewidth]{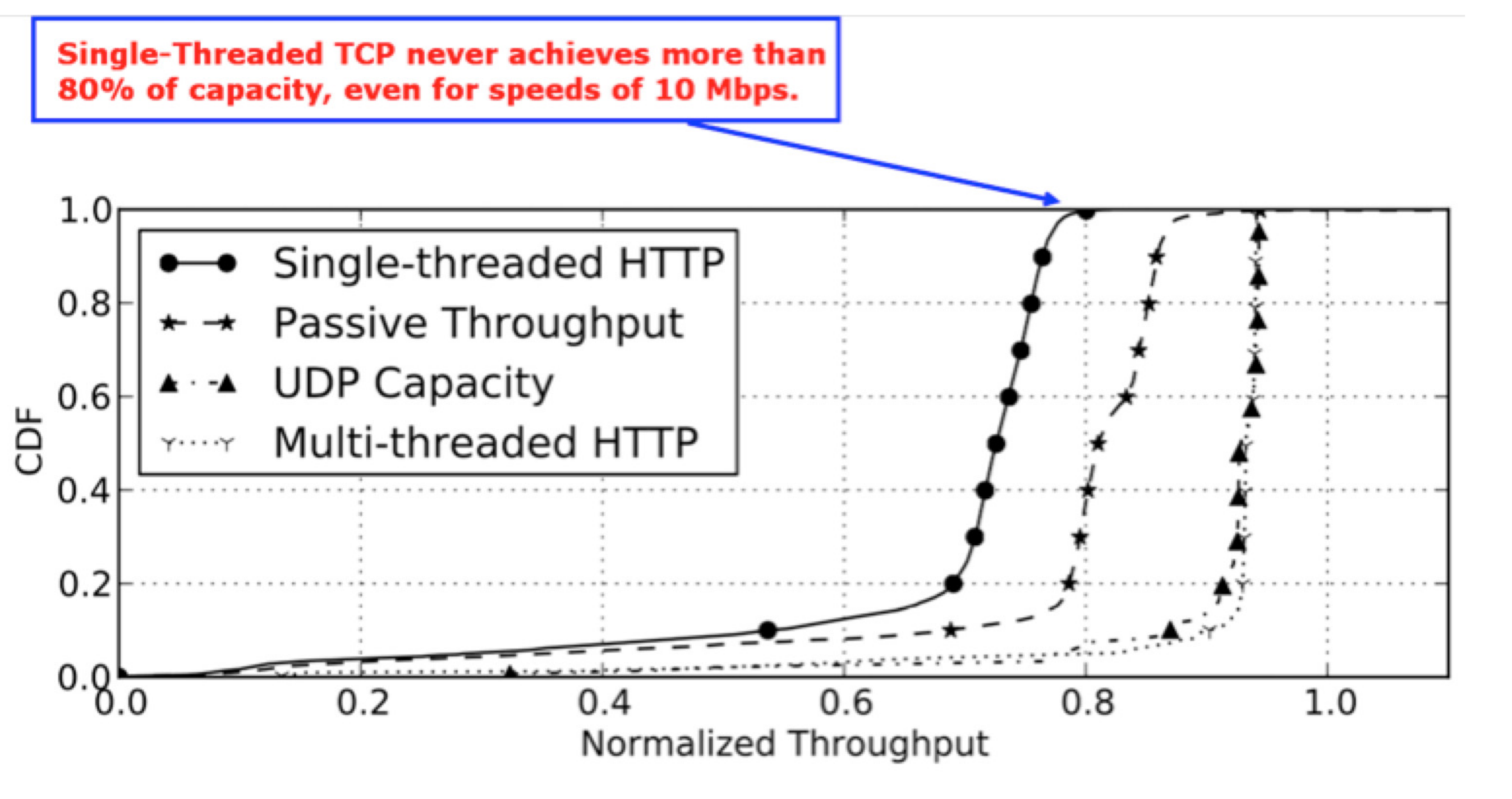}
    \caption{Throughput vs. number of TCP
    threads.\protect\cite{Sundaresan2011:bismark}}
\label{fig:throughput-numtcp}
\end{figure}

A single TCP connection cannot typically achieve a throughput approaching full
link capacity, for two reasons: (1) a single connection takes longer to send
at higher rates because TCP slow start takes longer to reach link capacity,
and (2) a single connection is more susceptible to temporarily slowing down
transmission rates when it experiences packet loss (a common occurrence on an
Internet path). 
Past research concluded that a
speed test should have at least four parallel connections to accurately
measure throughput~\cite{Sundaresan2011:bismark}.  For the same reason, modern web browsers typically open
as many as six parallel connections to a single server in order to maximize
use of available network capacity between the client and web server.

\paragraph{Test duration.} The length of a test and the amount of data
transferred also significantly affect test results. As previously described, a
TCP sender does not immediately begin sending traffic at full capacity but
instead begins in TCP slow start until the sending rate reaches a
pre-configured threshold value, at which point it begins AIMD congestion
avoidance.  As a result, if a transfer is too short, a TCP sender will spend a
significant fraction of the total transfer in TCP slow start, ensuring that
the transfer rate will fall far short of available capacity.  As
access speeds increase, most test tools have also needed to increase test
duration.

\paragraph{Throughput calculation.} The method that tests use to
calculate results appears to vary widely; often this method is not
disclosed. Tests may discard some high and/or low results, may use the median
or the mean, may take only the highest result and discard the rest, etc.
This makes different tests difficult to compare. Finally, some tests may
include all of the many phases of a TCP transfer, even though some of those
phases are necessarily at rates below the capacity of a link:

\begin{itemize}
	\itemsep=-1pt
\item	the slow start phase at the beginning of a transfer (which occurs in every
TCP connection); 
\item	the initial “additive increase” phase of the TCP transfer when the sender
is actively increasing its sending rate but before it experiences the first packet loss that results in multiplicative decrease;
\item any packet loss episode which results in a TCP timeout, and subsequent re-entry
into slow start
\end{itemize}

Estimating the throughput of the link is not as simple as dividing the amount
of data transferred by the total time elapsed over the course of the transfer.
A more accurate estimate of the transfer rate would instead measure the
transfer during steady-state AIMD, excluding the initial slow start period.
Many standard throughput tests, including the FCC/SamKnows test, omit the
initial slow start period. The Ookla test implicitly omits this period by
discarding low-throughput samples from its average measurement. Tests that
include this period will result in a lower value of average
throughput than the link capacity can support in steady state. 

\paragraph{Self-selection bias.} Speed tests that are initiated by a user
suffer from self-selection bias~\cite{heckman1990selection}:
many users initiate such tests only when they are experiencing a technical
problem or are reconfiguring their network.  For example, when configuring a
home wireless network, a user may run a test over WiFi, then re-position their
WiFi AP and run the test again.  These measurements may help the user optimize
the placement of the wireless access point but, by design, they reflect the
performance of the user's home wireless network, not that of the ISP.  Tests
that are user-initiated (``crowdsourced'') are more
likely to suffer from self-selection bias.
It can be difficult to use these results to draw conclusions about
an ISP, geographic region, and so forth. 

\paragraph{Infrequent testing.} If tests are too infrequent or are only taken
at certain times of day, the resulting measurements may not accurately reflect
a user's Internet capacity.  An analogy would be looking out a window once per
day in the evening, seeing it was dark outside, and concluding that it must be
dark 24 hours a day.  Additionally, if the user only conducts a test when
there is a transient problem, the resulting measurement may not be
representative of the performance that a user typically experiences.
Automatic tests run multiple times per day at randomly selected times during
peak and off-peak times can account for some of these factors.  

%% file: future.tex
\section{The Future of Speed Testing}\label{sec:future}

Speed testing tools will need to evolve as end user connections approach and
exceed 1~Gbps, especially given that so many policy, regulatory, and
investment decisions are based on speed measurements. As access network speeds
increase and the performance bottlenecks move elsewhere on the path, speed
test design must evolve to keep pace with both faster network technology and
evolving user expectations. We recommend the following:

\paragraph{Retire outmoded tools such as NDT.} NDT, also known as the Internet
Health Test~\cite{iht}, may appear at first glance to be suitable for speed tests. This
is not the case, though it continues to be used for speed measurement despite
its unsuitability and proven inaccuracy~\cite{feamster-fcc}. Its inadequacy for measuring access
link speeds has been well-documented~\cite{bauer}.  One significant problem is that NDT
still uses a single TCP connection, nearly two decades after this was shown to
be inadequate for measuring link capacity.  NDT is also incapable of reliably
measuring access link throughput for speeds of 100~Mbps or more, as we enter
an era of gigabit speeds.  The test also includes the initial TCP slow start
period in the result, leading to a lower value of average throughput than the
link capacity can support in TCP steady state.  It also faces all of the
user-related considerations that we discussed in
Section~\ref{sec:limitations}. It is time to retire the use of NDT for speed
testing and look ahead to better methods.

\paragraph{Use native, embedded, and dedicated measurement techniques and
devices.} Web-based tests (many of which rely on Javascript) cannot transfer
data at rates that exceed several hundred megabits per second. As network
speeds increase, speed tests must be ``native'' applications or run on
embedded devices (\eg, home router, Roku, Eero, AppleTV) or otherwise
dedicated devices (\eg, Odroid, Raspberry Pi, SamKnows ``white box'', RIPE
Atlas probes).

\paragraph{Control for factors along the end-to-end path when analyzing
results.} Section~\ref{sec:limitations} outlined many factors that can affect
the results of a speed test other than the capacity of the ISP link---ranging
from cross-traffic in the home to server location and provisioning. As access
ISP speeds increase, these limiting factors become increasingly important, as
bottlenecks elsewhere along the end-to-end path become increasingly prevalent.

\paragraph{Measure to multiple destinations.} As access network speeds begin
to approach and exceed 1~Gbps, it can be difficult to identify a single
destination and end-to-end path that can support the capacity of the access
link. Looking ahead, it may make sense to perform active speed test
measurements to multiple destinations simultaneously, to mitigate the
possibility that any single destination or end-to-end network path becomes the
network bottleneck.

\paragraph{Augment active testing with application quality metrics.} In many
cases, a user's {\em experience} is not limited by the access network speed,
but rather the performance of a particular application (\eg, streaming video)
under the available network conditions. As previously mentioned, even the most
demanding streaming video applications require only tens of megabits per
second, yet user experience can still suffer as a result of application
performance glitches, such as changes in resolution or rebuffering. As access
network speeds increase, it will be important to monitor not just ``speed
testing'' but also to develop new methods that can monitor and infer quality
metrics for a variety of applications.

\paragraph{Adopt standard, open methods to facilitate better comparisons.} It
is currently very difficult to directly compare the results of different speed
tests, because the underlying methods and platforms are so different. Tools
that select the highest result of several sequential tests, or
the average of several, or the average of several tests after the highest and
lowest have been discarded. As the FCC has stated~\cite{fcc-miller}:
``A well documented, public methodology for tests is critical
to understanding measurement results.'' Furthermore, tests and networks should
disclose any circumstances that result in the prioritization of speed test
traffic.

Beyond being well-documented and public, the community should also come to
agreement on a set of standards for measuring access link performance and
adopt those standards across test implementations.

%% file: paper-arxiv.bbl
\begin{thebibliography}{10}

\bibitem{apple}
{Apple: Resolve Wi-Fi and Bluetooth Issues Caused by Wireless Interference},
  2019.
\newblock \url{https://support.apple.com/en-us/HT201542}.

\bibitem{bauer}
S.~Bauer, D.~D. Clark, and W.~Lehr.
\newblock {Understanding Broadband Speed Measurements}.
\newblock In {\em Technology Policy Research Conference (TPRC)}, 2010.

\bibitem{bufferbloat-1}
{Bufferbloat}.
\newblock \url{https://www.bufferbloat.net}.

\bibitem{calspeed}
{CALSPEED Program}, 2019.
\newblock \url{http://cpuc.ca.gov/General.aspx?id=1778}.

\bibitem{eetimes}
{Avoiding Interference in the 2.4-GHz ISM Band}, 2006.
\newblock \url{https://www.eetimes.com/document.asp?doc_id=1273359}.

\bibitem{fcc-aug-2013}
{FCC Re: Measuring Broadband America Program (Fixed), GN Docket No. 12­264},
  Aug. 2013.
\newblock \url{https://ecfsapi.fcc.gov/file/7520939594.pdf}.

\bibitem{fcc-mba}
{FCC: Measuring Broadband America Program}.
\newblock \url{https://www.fcc.gov/general/measuring-broadband-america}.

\bibitem{fcc-mba2011}
{Measuring Broadband America}.
\newblock Technical report, Federal Communications Commission, 2011.
\newblock
  \url{https://transition.fcc.gov/cgb/measuringbroadbandreport/Measuring_U.S._-_Main_Report_Full.pdf}.

\bibitem{mba-report-2014}
{FCC MBA Report 2014}, 2014.
\newblock
  \url{https://www.fcc.gov/reports-research/reports/measuring-broadband-america/measuring-broadband-america-2014}.

\bibitem{fcc-miller}
{M-Lab Discussion List: MLab speed test is incorrect?}, 2018.
\newblock
  \url{https://groups.google.com/a/measurementlab.net/forum/\#!topic/discuss/vOTs3rcbp38}.

\bibitem{feamster-fcc}
{Letter to FCC on Docket No. 17-108}, 2014.
\newblock
  \url{https://ecfsapi.fcc.gov/file/1083088362452/fcc-17-108-reply-aug2017.pdf}.

\bibitem{ftp-rate}
{FTP Rate Limiting}, 2019.
\newblock {\url{https://forum.filezilla-project.org/viewtopic.php?t=25895}}.

\bibitem{bufferbloat-2}
J.~Gettys.
\newblock {Bufferbloat: Dark Buffers in the Internet}.
\newblock In {\em IEEE Internet Computing}, 2011.

\bibitem{heckman1990selection}
J.~J. Heckman.
\newblock Selection bias and self-selection.
\newblock In {\em Econometrics}, pages 201--224. 1990.

\bibitem{iht}
{Internet Health Test}, 2019.
\newblock \url{http://internethealthtest.org/}.

\bibitem{lifewire}
{Change the Wi-Fi Channel Number to Avoid Interference}, 2018.
\newblock
  \url{https://www.lifewire.com/wifi-channel-number-change-to-avoid-interference-818208}.

\bibitem{apnic-blog}
J.~Livingood.
\newblock {Measurement Challenges in the Gigabit Era}, June 2018.
\newblock
  \url{https://blog.apnic.net/2018/06/21/measurement-challenges-in-the-gigabit-era/}.

\bibitem{minnestoa}
{CheckspeedMN}.
\newblock \url{https://mn.gov/deed/programs-services/broadband/checkspeedmn}.

\bibitem{newyork-1}
{A.G. Schneiderman Encourages New Yorkers To Test Internet Speeds And Submit
  Results As Part Of Ongoing Investigation Of Broadband Providers}, 2017.
\newblock
  \url{https://ag.ny.gov/press-release/ag-schneiderman-encourages-new-yorkers-test-internet-speeds-and-submit-results-part}.

\bibitem{newyork-2}
{Are You Getting The Internet Speeds You Are Paying For?}
\newblock \url{https://ag.ny.gov/SpeedTest}.

\bibitem{newyork-3}
{New York State Broadband Program Office - Speed Test}.
\newblock \url{https://nysbroadband.ny.gov/speed-test}.

\bibitem{nginx-rate}
{Nginx Rate Limiting}, 2019.
\newblock {\url{https://www.nginx.com/blog/rate-limiting-nginx/.}}

\bibitem{ofcom}
{Broadband Speeds: Research on fixed line home broadband speeds, mobile
  broadband performance, and related research}.
\newblock
  \url{https://www.ofcom.org.uk/research-and-data/telecoms-research/broadband-research/broadband-speeds}.

\bibitem{oracle-pricing}
{Oracle IaaS Pricing}, 2019.
\newblock {\url{https://cloud.oracle.com/en\_US/iaas/pricing.}}

\bibitem{pc-world}
{Six Things That Block Your Wi-Fi, and How to Fix Them}, 2011.
\newblock
  \url{https://www.pcworld.com/article/227973/six_things_that_block_your_wifi_and_how_to_fix_them.html}.

\bibitem{pennsylvania}
{A Broadband Challenge: Reliable broadband internet access remains elusive
  across Pennsylvania, and a Penn State faculty member is studying the issue
  and its impact}, 2018.
\newblock
  \url{https://news.psu.edu/story/525994/2018/06/28/research/broadband-challenge}.

\bibitem{revolution-wifi}
{The 2.4 GHz Spectrum Congestion Problem and AP Form-Factors}, 2015.
\newblock
  \url{http://www.revolutionwifi.net/revolutionwifi/2015/4/the-dual-radio-ap-form-factor-is-to-blame-for-24-ghz-spectrum-congestion}.

\bibitem{samknows-method}
{SamKnows Test Methodology White Paper}, Dec. 2011.
\newblock
  \url{https://availability.samknows.com/broadband/uploads/Methodology_White_Paper_20111206.pdf}.

\bibitem{Sundaresan2011:bismark}
S.~Sundaresan, W.~De~Donato, N.~Feamster, R.~Teixeira, S.~Crawford, and
  A.~Pescap{\`e}.
\newblock {Broadband Internet Performance: A View from the Gateway}.
\newblock In {\em ACM SIGCOMM}, pages 134--145, Aug. 2011.

\bibitem{wtf}
S.~Sundaresan, N.~Feamster, and R.~Teixeira.
\newblock {Home Network or Access Link? Locating Last-mile Downstream
  Throughput Bottlenecks}.
\newblock In {\em International Conference on Passive and Active Network
  Measurement (PAM)}, pages 111--123, 2016.

\bibitem{wehe}
{Wehe}, 2019.
\newblock {\url{https://dd.meddle.mobi}}.

\bibitem{xfinity-requirements}
{Xfinity Internet Minimum System Recommendations}.
\newblock
  \url{https://www.xfinity.com/support/articles/requirements-to-run-xfinity-internet-service}.

\end{thebibliography}
